\documentstyle[12pt]{article}
\begin{document}

\tolerance=5000

\def\pp{{\, \mid \hskip -1.5mm =}}
\def\cL{{\cal L}}
\def\be{\begin{equation}}
\def\ee{\end{equation}}
\def\bea{\begin{eqnarray}}
\def\eea{\end{eqnarray}}
\def\tr{{\rm tr}\, }
\def\nn{\nonumber \\}
\def\e{{\rm e}}
\def\D{{D \hskip -3mm /\,}}
\def\la{\label}
\def\PLB#1 {\Jl{Phys. Lett.}{#1B}}

\ \hfill
\begin{minipage}{3.5cm}
\end{minipage}


\begin{center}
{\Large\bf Accelerating cosmology in modified gravity with scalar field}

\vspace{6mm}

{\sc Yulia A. Shaido}$^{\dagger\clubsuit}$\footnote{E-mail: Shaido@ngs.ru} 
and {\sc Akio Sugamoto}$^{\ddagger}$\footnote{E-mail: sugamoto@phys.ocha.ac.jp} \vspace{3mm}

{\sl $^\dagger$Laboratory for Fundamental Studies, 
\\
Tomsk State Pedagogical University,
\\
634041 Tomsk, RUSSIA}
\vspace{3mm}

{\sl $^\clubsuit$Dept. of Physics and Chemistry, Graduate School of Humanities and Sciences, Ochanomizu University, JAPAN} 
\vspace{3mm}

{\sl $^{\ddagger}$Dept. of Physics, Faculty of Science, Ochanomizu University,
\\
1-1 Otsuka 2, Bunkyo-ku, Tokyo 112-8610, JAPAN}

\vfill

{\bf Abstract}
\end{center}

The modified gravity with $1/R$ term (R being scalar curvature) and the
Einstein-Hilbert
term is studied by incorporating the phantom scalar field. A number of
cosmological solutions are derived in the presence of the phantom field in the
 perfect fluid background. It is shown the current inflation obtained in
the modified gravity is affected by the existence the phantom field. 

 \vfill

\noindent 

\newpage

It became clear recently that our current universe is accelerating.
There are various scenarios to explain such current acceleration.
One of the attractive scenarios is to modify the gravitational dynamics 
in such a way that with decrease of gravity, the effective gravitational
dark energy appears. 
The interesting model of that sort has been considered in ref.\cite{CHT}
by modifying of Einstein gravity with 1/R term.
It is interesting that such correction maybe induced by M-theory \cite{SN}. 
Unfortunately, such theory
is inconsistent and contains number of instabilities \cite{DKCh}. Nevertheless, the
consistent modification of such theory maybe done by adding of higher
derivative terms which are responsible also for early time inflation.
It was shown in refs.\cite{SN1} that such modified gravity is viable and
may describe both early time inflation and current acceleration.
Moreover, such higher derivative terms which make theory stable at low
curvature maybe induced by conformal anomaly too, see explicit example in
ref.\cite{SN1, ABF}. It is appealing to consider the role of other types of
matter (in particulary, of other types of dark energy) to such modified
gravity. Being in the phase with low gravity one may effectively forget
about higher derivative terms and consider the simplest modified gravity.

We focus on the simplest correction to the Einstein-Hilbert action, by introducing $R^{-1}$ term \cite{CHT}
\\
\begin{equation}
S=\frac{M^2_P}{2}\int{d^4x\sqrt{-g}\bigg({R-\frac{\mu^4}{R}}\bigg)}+\int{d^4x\sqrt{-g}(L+L_{fluid})}.
\end{equation}
\\
Here $\mu$ is a new parameter with the unit of [mass], ${M_P\equiv{(8\pi{G})^{-\frac{1}{2}}}}$ is the Planck mass and $L$ is the Lagrangian density for matter: 
\\
\begin{equation}
L=-\frac{\epsilon}{2}g^{\mu\nu}\partial_{\mu}y\partial_{\nu}y,
\end{equation}
\\
where we have introduced a signature $\epsilon$. If $\epsilon>0$, 
$y$ is the ordinary scalar field, but if $\epsilon<0$, 
$y$ is the phantom field. The phantom field appearence is not clear but it
has many similarities with quantum field theory \cite{negative}. 
The field equation for the metric is then
\\
\begin{equation}
\bigg({1+\frac{\mu^4}{R^2}}\bigg){R_{\mu\nu}}-\frac{1}{2}\bigg({1-\frac{\mu^4}{R^2}}\bigg){Rg_{\mu\nu}}
+\mu^4{\bigg({g_{\mu\nu}{\nabla_{\alpha}}{\nabla^{\alpha}}-\nabla_{(\mu}\nabla_{\nu)}}\bigg)}{R^{-2}}
=\frac{T_{\mu\nu}}{M^2_P},
\end{equation}
\\
where $T_{\mu \nu }$ is the energy-momentum tensor of the matter, including also the contribution from a perfect fluid.
\\
In the following we consider the case of phantom field ($\epsilon<0$).
Then, we have 
\\
\begin{equation}
T_{\mu\nu}=(\rho+P){\cup_{\mu}}{\cup_{\nu}}+Pg_{\mu\nu}-{\partial_{\mu}}y{\partial_{\nu}}y
+\frac{1}{2}g_{\mu\nu}g^{\alpha\beta}{\partial_{\alpha}}y{\partial_{\beta}}y,
\end{equation}
\\
where $\rho$ is the energy density, $P$ is the pressure and $\cup^{\alpha}$ is the velocity of the fluid. 
\\
We take the metric to be of the flat Robertson-Walker form, giving
\\
\begin{equation}
ds^2=-dt^2+a^2(t)dx^2
\end{equation}
\\
for which the scalar curvature satisfies
\\
\begin{equation}
R=6(\dot {H}+2H^2),
\end{equation}
\\
where an overdot denotes differentiation with respect to time, $H=\frac{\dot {a}}{a}$ and $a(t)$ is the scale factor.
\\ 
The time-time component of the field equations for this metric reads
\\
\begin{equation}
3H^2-\frac{\mu^4}{12({\dot{H}}+2H^2)^3}\bigg({2H\ddot{H}+15H^2{\dot{H}}+2\dot{H}^2+6H^4}\bigg)=\frac{\rho-\frac{1}{2}{\dot{y}}^2}{M^2_P}.
\end{equation}
\\
\\
The space-space component of the field equations for this metric is,
\\
\begin{equation}
\dot {H}+\frac{3}{2}H^2-\frac{\mu ^4}{2R^2}\bigg[{4\dot {H}+9H^2-R^2\partial _{0}\partial _{0}R^{-2}-2HR^2\partial _{0}R^{-2}}\bigg]=-\frac{2P-\dot{y}^2}{4M^2_P}.
\end{equation}
\\
\\
We are going to take the simplest case of $y(t)=ct+b$ \cite{Gibbons}, where
$c$ and $b$
are constants. This is a solution of the field equation for $y$. Then, we obtain the following equations
\\
\begin{equation}
3H^2-\frac{\mu^4}{12({\dot{H}}+2H^2)^3}\bigg({2H\ddot{H}+15H^2{\dot{H}}+2\dot{H}^2+6H^4}\bigg)=\frac{\rho-\frac{1}{2}c^2}{M^2_P},
\end{equation}
\\
\begin{equation}
\dot {H}+\frac{3}{2}H^2-\frac{\mu ^4}{2R^2}\bigg[{4\dot {H}+9H^2-R^2\partial _{0}\partial _{0}R^{-2}-2HR^2\partial _{0}R^{-2}}\bigg]=-\frac{2P-c^2}{4M^2_P}.
\end{equation}
\\
Having these equations, we will study the evolution of our universe in the Einstein-Hilbert theory modified by $R^{-1}$, in the presence of the phantom field and the perfect fluid background.
\\
However, as we see these equations are too complicated to be solved analytically, 
so that it is necessery to look for their solutions approximately. 

First let us look for the following Ansatz:
\\
\begin{equation}
a(t)=ke^{\bar{H}t}
\end{equation}
\\
where $\bar{H}$ and $k$ are positive constants.
\\
Substituting this ansatz (11) into (9) and (10), and considering that 
$\rho(t)$ and $P(t)$ decrease exponentially with this ansatz,
 we obtain at a sufficiently late time 
\\
\\
\begin{equation}
\bar{H}^2=1-\frac{c^2}{3\mu^4M^2_P}.
\end{equation}
\\
If $c^2<3{\mu^4}M^2_P$, then we have the de Sitter expansion of our universe. 

Next, we make the following conformal transformation from the fields \cite{CHT} and variables without tilde to those with tilde: 
$\tilde {g}_{\mu \nu }=p(\phi)g_{\mu \nu }$, $d\tilde t=\sqrt{p}dt$ and $\tilde {a}(t)=\sqrt{p}a(t)$, 
where $p\equiv {exp\bigg({\sqrt{\frac{2}{3}}\frac{\phi }{M_P}}\bigg)}\equiv1+\frac{\mu^4}{R^2}$ and $\phi$ is a real scalar function on space-time. 
\\
Then, the energy momentum tensor in the new frame (Einstein frame) 
described in terms of fields and variables with tilde, is given by
\\
\begin{equation}
\tilde{T}_{\mu \nu }= \bigg({\tilde{\rho}
+\tilde {P}}\bigg)\tilde{\cup}_{\mu }\tilde{\cup }_{\nu }
+\tilde {P}\tilde{g}_{\mu \nu }
-\tilde{\partial }_{\mu }\tilde{y}\tilde{\partial }_{\nu }\tilde{y}
+\frac{1}{2}\tilde {g}_{\mu \nu }\tilde{g}^{\alpha \beta }
\tilde{\partial }_{\alpha }\tilde{y}\tilde{\partial }_{\beta }\tilde {y}, 
\end{equation}
where $\tilde{\cup}_{\alpha }\equiv {\sqrt{p}\cup_{\alpha }}$, 
$\tilde{\rho }=\frac{\rho}{p^2}$, $\tilde{P}=\frac{P}{p^2}$, $\tilde{\partial}_{\lambda}=\sqrt{p}{\partial}_{\lambda}$ and $\tilde{y}=\frac{1}{p}y$. 
\\
\\
In terms of the new metric $\tilde{g}_{\mu \nu }$, our theory is that of 
a scalar field $\phi(x^{\mu })$ minimally coupled to Einstein gravity, 
and non-minimally coupled to matter, with potential
\\
\begin{equation}
V(\phi)=\mu^2M^2_P\frac{\sqrt{p-1}}{p^2}.
\end{equation} 
\\
\\
Note, that generally speaking, gravitational physics does not depend on
the frame choice \cite{tkach} while it is often more convenient to
work in a specific frame. 
Denoting all quantities (except $\phi$) in the Einstein frame with tilde, 
the relevant Einstein-frame cosmological equations of motion are
\\
\\
\begin{equation}
3\tilde{H}^2=\frac{1}{M^2_P}{\bigg[{\rho_{\phi}+\tilde{\rho}+\tilde{\rho}_y}\bigg]}, 
\end{equation} 
\\
\\
\begin{equation}
\ddot{\phi}+3\tilde{H}\dot{\phi}+\frac{dV}{d\phi}(\phi)-\frac{1}{\sqrt{6}}\bigg[{\tilde{\rho}-3\tilde{P}+\tilde{\rho}_y-3\tilde{P}_{y}}\bigg]=0, 
\end{equation} 
\\
\\
where
\\
\begin{equation}
\rho_{\phi}=\frac{1}{2}\dot{\phi}^2+V(\phi),
\end{equation} 
\\
\begin{equation}
\tilde{\rho}_y=-\frac{1}{2p^2}\dot{\tilde{y}^2},
\end{equation} 
\\
\begin{equation}
\tilde{\rho}=\frac{C}{\tilde{a}^{3(1+\omega)}}exp{\bigg[{-\frac{(1-3\omega)}{\sqrt{6}}\frac{\phi}{M_P}}\bigg]},
\end{equation} 
with constants $C$ and $\omega$. Here $\omega$ is defined by $\omega=\frac{P}{\rho}$.
\\
\\
The Hubble parameter $H$ in the matter-frame is related to that in the Einstein frame $\tilde{H}\equiv{\frac{\dot{\tilde{a}}}{\tilde{a}}}$ by
\\
\begin{equation}
H=\sqrt{p}{\bigg({\tilde{H}-\frac{\dot{\phi}}{M_P\sqrt{6}}}\bigg)}.
\end{equation} 

Now let us first focus on vacuum cosmological solutions, when $\tilde{\rho}=\tilde{P}=0$, $\tilde{\rho}_{y}=\tilde{P}_{y}=0$.
Furthermore, we are going to take the simplest case when the potential is well-approximated by
\\
\begin{equation}
V(\phi)\simeq{\mu^2M^2_Pexp{\bigg({-\sqrt{\frac{3}{2}}\frac{\phi}{M_P}}\bigg)}}.
\end{equation}
\\
Then, we may obtain the equations of motion:
\\
\begin{equation}
3\tilde{H}^2{\simeq}\frac{1}{M^2_P}\bigg[{\frac{1}{2}\dot{\phi}^2+\mu^2{M^2_P}exp{\bigg(-\sqrt{\frac{3}{2}}\frac{\phi}{M_P}}\bigg)}\bigg], 
\end{equation}
\\
\begin{equation}
\ddot{\phi}+3\tilde{H}\dot{\phi}-\sqrt{\frac{3}{2}}\mu^2{M_P}{exp{\bigg({-\sqrt{\frac{3}{2}}\frac{\phi}{M_P}}\bigg)}}{\simeq}0. 
\end{equation}
\\
Looking for solutions of the form
\begin{equation}
\tilde{a}(\tilde{t}){\simeq}\tilde{t}^\alpha,
\end{equation}
\begin{equation}
\phi{\simeq}B\ln{\tilde{t}}.
\end{equation}
\\
 we obtain
\\
\begin{equation}
9\alpha^2{\simeq}4+3\mu^2
\end{equation}
and
\begin{equation}
-4+12\alpha-3\mu^2{\simeq}0,
\end{equation}
what gives $\alpha{\simeq}\frac{4}{3}$, $\mu\simeq2$ for $M_P=1$.
This means that the universe is in the phase of late time inflation. We call this the solution $(A)$.

Then we introduce the phantom field $\tilde{y}=k_1\tilde{t}+d_1$, but still keep $\tilde{\rho}=0$ and $V(\phi)=0$. The Eqs. (22) and (23) become
\\
\begin{equation}
3\frac{\alpha^2_1}{\tilde{t}^2}{\simeq}\frac{B^2_1}{2\tilde{t}^2}-\frac{k^2_1}{2\tilde{t}^{2\sqrt{\frac{2}{3}}B_1}},
\end{equation}
\\
\begin{equation}
-\frac{B_1}{\tilde{t}^2}+3\frac{\alpha_1B_1}{\tilde{t}^2}-\frac{k^2_1}{\sqrt{6}\tilde{t}^{2\sqrt{\frac{2}{3}}B_1}}{\simeq}0.
\end{equation}
\\
From these we have $B_1=\sqrt{\frac{3}{2}}$. Then, we obtain the following equation for $\alpha_1$ $(>0)$,
\\
\begin{equation}
4\alpha^2_1+6\alpha_1-3\simeq0,
\end{equation}
\\
which gives
\\
\begin{equation}
\alpha_1{\simeq}\frac{2}{5},
\end{equation}
\\
and
\\
\begin{equation}
k^2_1{\simeq}\frac{3}{5}.
\end{equation}
\\
This is not a good solution, since the universe does not expand and shrinks as $a(t){\simeq}t^{-\frac{1}{5}}$.

Therefore, we consider the case when $\tilde{\rho}\neq{0}$, having $V(\phi)=0$, $\tilde{a}(\tilde{t}){\simeq}\tilde{t}^{\alpha_2}$, $\phi{\simeq}B_2\ln{\tilde{t}}$ and $\tilde{y}=k_2\tilde{t}+d_2$. 
Then $\alpha_2$ and $\omega$ satisfies the following equation
\\
\begin{equation}
(6\alpha_2-3)(1+\omega)=0,
\end{equation}
\\
for $B_2=\sqrt{\frac{3}{2}}$. The solution is $\alpha_2=\frac{1}{2}$ or $\omega=-1$.
If $\alpha_2=\frac{1}{2}$, then $\omega=\frac{1}{k^2_2}-\frac{1}{3}$ holds. We call this the solution $(B)$. On the other hand if $\omega=-1$, then $\alpha_2=\frac{3{\pm}\sqrt{9-16k^2_2}}{8}$ holds. This solution is, however, not interesting, since the universe does not expand and shrinks. 

Furthermore we consider the case when $\tilde{\rho}\neq{0}$ and $V(\phi)$ is
 non-vanishing in the form (21). If we assume that
 $\tilde{a}(\tilde{t}){\simeq}\tilde{t}^{\alpha_3}$, $\phi{\simeq}B_3\ln{\tilde{t}}$ and $\tilde{y}=k_3\tilde{t}+d_3$, 
then we have for $\alpha_3$ and $\omega$ the equivalent solutions as
 in the foregoing case, when $B_3=\sqrt{\frac{3}{2}}$. The solution is
 also $\alpha_3=\frac{1}{2}$ or $\omega=-1$. Now, if $\alpha_3=\frac{1}{2}$,
 then $\omega=\frac{5k^2_3}{6C}-\frac{1}{2C}-\frac{2}{3}$.
 This is solution $(B')$ depending on the parameter $C$. 
If $\omega=-1$, then $\alpha _3=\frac{1{\pm}\sqrt{1-4D}}{2}$, 
where $D=\frac{C}{9}+\frac{5}{18}k^2_3+\frac{1}{12}>\frac{1}{4}$. 
This solution is of limited physical interest.

Now we will give the physical discussion. Let us consider the most 
interesting cases $(A)$, $(B)$ and $(B')$. First we consider the solution
 $(A)$. In this solution $\alpha{\simeq}\frac{4}{3}$, that is
 $\omega_{eff}=-\frac{2}{3}$. Then, the universe evolves following the
 late-time power-law inflation. In terms of physical time it gives
 $a(t)=t^2$. Hence, the second derivative of $a(t)$ with respect to time
 is positive, so one has the acceleration cosmology at late time.
After introducing the phantom field, we have solutions $(B)$ and $(B')$.
 In case of the solution $(B)$, $\alpha_2=\frac{1}{2}$, and
 then $\omega_{eff}=\frac{1}{k^2_2}-\frac{1}{3}$. 
We take $k$ as a positive constant. If $k_2{\rightarrow}\infty$, then
$\omega_{eff}$ tends to $-\frac{1}{3}$. Now, in this case we receive that
in
matter frame $a(t)$ becomes constant. In the solution $(B')$, $\alpha_3=\frac{1}{2}$, but $\omega_{eff}$ depends on $C$ and $k_3$. Therefore, we can receive different physical results from the solution.

There are several interesting directions where our study maybe
generalized.
First of all, modified gravity after transition to Einstein frame looks
like usual General Relativity with extra scalar (scalar-tensor or
dilatonic gravity).
It would be interesting to investigate the role of such scalar-tensor
background to energy
conditions in phantom cosmology \cite{negative, Gibbons}.
Second, as transformed theory looks like dilatonic gravity,
it would be of interest to study the role of quantum dilaton effects to
our cosmology, for instance, in the way discussed in 
\cite{DCA} via account of dilaton dependent four-dimensional trace
anomaly (for a review, see \cite{review}).
This will be discussed elsewhere.
\vspace{3mm}
\\
\\
\noindent {\bf Acknowledgments}

We are grateful to S.D. Odintsov for stimulating discussions.
One of the authors (Y.S.) is partially supported by Association of International Education, Japan (AIEJ) Short-term Student Exchange Promotion Program Scholarship.
\\

\end{document}